\documentclass[%
 reprint,
 amsmath,amssymb,
 aps,
prb,
]{revtex4-2}

  \usepackage[utf8]{inputenc}                  
  \usepackage[english]{babel}                  
  \usepackage[T1]{fontenc}                     
  \usepackage{lmodern}                         
  \usepackage{dcolumn}                         
  \usepackage{graphicx}                        
  \usepackage{floatflt}                        
  \usepackage{placeins}  					
  \usepackage{xcolor}                          
  \usepackage{amsmath}
  \usepackage{amsfonts}
  \usepackage{amssymb}
  \usepackage{mathtools}
  \usepackage{bm}
  \usepackage{nicefrac}
  \usepackage{dsfont}
  \usepackage{mathrsfs}
  \usepackage{multirow}                        
  \usepackage{tabularx}                        
  \usepackage{longtable}                       
  \usepackage[breaklinks=true]{hyperref}
  \usepackage{cleveref}
  \usepackage[]{siunitx} 

\usepackage{filecontents}

\newcommand{\abs}[1]{\ensuremath{ \left| #1 \right| }}                    
\newcommand{\mean}[1]{\ensuremath{ \left\langle#1\right\rangle }}         

\newcommand{\imag}[1]{\operatorname{Im}\left( #1 \right)}

\newcommand{\dd}[0]{\mathrm d}                
\newcommand{\const}[0]{\operatorname{const.}} 
\newcommand{\del}[0]{\partial}                

\renewcommand{\rho}[0]{\varrho}
\renewcommand{\theta}[0]{\vartheta}
\renewcommand{\phi}[0]{\varphi}

\renewcommand{\vec}[1]{\bm{#1}}        

\newcommand{\ba}[1]{\begin{align} #1 \end{align}}

\newcommand{\muS}{\mu_s}
\newcommand{\kB}{k_\mathrm{B}}
\newcommand{\Neel}{N\'{e}el}

\begin{document}

\preprint{APS/123-QED}

\title{Spin-noise spectroscopy as a tool for probing magnetic order}

\author{Julius Schlegel}%
  \email{julius.schlegel@uni-konstanz.de}
\author{Martin Evers}%
\author{Ulrich Nowak}
\affiliation{%
  University of Konstanz
}%

\date{\today}

\begin{abstract}
Spin noise spectroscopy is a technique to measure magnetization fluctuations, a subject of increasing relevance in ultrafast spintronics.   We investigate numerically the equilibrium spin noise of ferro- and antiferromagnets within an atomistic spin model.
  The aim is to predict the possible outcomes of ultrafast spin-noise spectroscopy measurements and demonstrate  what relevant information can be extracted.
  Specifically, we show how this method can be used to determine phase transitions, frequencies of magnon modes and correlation times.
\end{abstract}

\maketitle

\section{Introduction}

Probing the magnetic order of antiferromagnets and their dynamics can be surprisingly difficult because of their lack of a global magnetization and its corresponding stray field. This is especially true for thin films and nanostructures. 
A recent topic of interest, for example, is spin transport through multilayer systems \cite{Wang14_AFMSpinTransportYIGintoNiO,Wang15_FMRexcitationOfAFMinsulators,Qiu2016_SpinCurrentProbeForAFMPhaseTransitions,Cramer18_SpinValve,Schlitz18_SpinHallMagnetoresist_AFMBilayer_NeelTemp}.
Here, the magnetic order is often not obvious as interface effects can play role leading to \SI{90}{\degree} coupling between ferromagnetic and adjacent antiferromagnetic layers. For thin layers, however,  methods like neutron scattering may not be feasible for the detection of their magnetic order \cite{Nemec18_OpticalProbingAFMOrder}.
Hence, novel experimental methods are are sought for and in this work we discuss the possibility to utilize ultrafast spin-noise spectroscopy for this purpose \cite{Starosielec08_UltrafastSpinNoiseSpetroscopy}.

Spin-noise spectroscopy \cite{Aleksandrov81_SpinNoiseSpectroscopy,Zapasskii13_ReviewSpinNoiseSpectroscopy,Sinitsyn16_TheorySpinNoiseSpectroscopy} is a technique to measure the fluctuations of the magnetization $m(t)$
by virtue of correlation functions. Typically, one is interested in the autocorrelation function $\mathrm{Corr}(\Delta t) = \mean{m(t)m(t+\Delta t)}$, but also higher order correlators can be of interest \cite{Li13_HigherOrderSpinNoise}.
The magnetization signal is measured by optical means, typically using Faraday rotation of laser light---which measures light transmitted through a sample \cite{Weiss23_SpinNoise_Orthoferrite}. 
But also a Kerr microscopy setup can be utilized---where the reflected light is detected \cite{Poltavtsev14_SpinNoiseKerr}.
Optical spectroscopy is not the only way to probe magnetic noise, also electrical detection is possible as carried out, for instance, on magnetic recording media \cite{Silva90_MagneticNoiseRecordingMedia}, however, in such setups the accessible frequencies are rather limited.
In particular there is no way to reach the relevant frequencies of antiferromagnets, which are typically in the terahertz regime.
In the past the focus of optical spin-noise spectroscopy has been on atoms and molecules \cite{Crooker04_SpinNoiseSpectroscopyAlkaliAtoms}.
The study of solid state systems is more challenging since here the strength of the fluctuations --- measured via the variance $\mean{m^2} - \mean{m}^2$ --- scales with $\nicefrac{1}{N}$, with $N$ being the number of spins.
Hence, the signal is supposed to be very small for macroscopic samples, and should vanish in the thermodynamic limit. 
The study of spin noise in semiconductors has a successful history \cite{Roemer07_SpinNoiseSpectroscopySemiconductors}, since the number of spins per volume is still small compared to magnetically ordered solids.
But improved sensitivity allows nowadays also to study the latter \cite{Balk18_NoiseSpectroscopyFerroSpinReorientation}, and further possibilities remain to be explored.
Promising candidates are especially antiferromagnets, where the average magnetization $\mean{m}$ is zero or very small in canted configurations and may not be probed directly. However, it has been shown that measuring magnetization  fluctuations can unveil unexpected material properties in an antiferromagnet, like an ultrafast random telegraph noise \cite{Weiss23_SpinNoise_Orthoferrite}.

In this work we explore possibilities of using optical spin-noise spectroscopy in magnetically ordered solids from a theoretical point of view.
The focus lies especially on the determination of phase transitions, the (anti)ferromagnetic-paramagnetic phase transition as well as an antiferromagnetic reorientation transition, the spin flop.
At phase transitions the noise amplitude usually peaks (or even diverges in case of ferromagnets), an advantage for experiments in terms of large signals.
This paper is organized as follows: we introduce our atomistic spin model used for the numerical investigations and explain how we calculate the spectral noise power density and the autocorrelation function, both properties which could be measured in an experiment.
Then we study two model systems, an easy-axis ferromagnet and an easy-axis antiferromagnet in an external field.
We identify several characteristics which can be used to probe the magnetic order in general and phase transitions in particular.

%

\section{Atomistic spin model}  \label{sec:ASM}
Our study is based on a classical, atomistic spin model \cite{Nowak07_SpinModels}, comprising $N$ Heisenberg spins, i. e.\ normalized magnetic moments $\vec{S}^l = \vec{\mu}^l/\muS$, which can point in any three-dimensional spatial direction.
These spins are positioned on regular lattice sites $\vec{r}^l$, where we assume a simple cubic lattice with lattice constant $a$, which is of size $N = N_x \times N_y \times N_z$.
The spin Hamiltonian reads
\ba{
  H = & - \sum_{n=1}^N \sum_{ \substack{m \in \\ \operatorname{NN}(n)}} \frac{J}{2} \vec{S}^n \cdot \vec{S}^m - d_z\sum_{n=1}^N \left(S_z^n\right)^2 \nonumber \\
      & - \muS\sum_{n=1}^N\vec{S}^n\cdot\vec{B},  \label{eq:Hamiltonian}
}
taking into account the Heisenberg nearest neighbors exchange interaction $J$, where the sum over $m$ runs over the $N_\mathrm{nb}$ nearest neighbors of $n$---counting each pair interaction twice in the double sum.
Furthermore, a uniaxial anisotropy with respect to the $z$ direction with anisotropy constant $d_z$ is included as well as a magnetic field $\vec{B}$. Along all three directions periodic boundary conditions are used.

The time evolution for each spin variable is given by the Landau-Lifshitz-Gilbert (LLG) equation \cite{Landau35_LL_equation,Gilbert55_Gilbert_damp,Gilbert04_Gilbert_damp_IEEE}, 
\ba{
  \frac{\dd \vec{S}^l}{\dd t} & = -\frac{\gamma}{\muS(1 + \alpha^2)}\Big[ \vec{S}^l \times \left( \vec{H}^l + \alpha\vec{S}^l \times \vec{H}^l \right) \Big]  \label{eq:LLG} \\
           \mbox{with}  & \quad       \vec{H}^l  = -\frac{\del H}{\del \vec{S}^l} + \vec{\xi}^l,  \nonumber
}
describing the motion of each spin in its effective field $\vec{H}^l$. Here, $\gamma$ is the absolute value of the gyromagnetic ratio and $\alpha$ the Gilbert damping constant.
The effective field comprises a deterministic part given by the derivative of the Hamiltonian and a stochastic part---the thermal noise $\vec{\xi}$, modeling a coupling of the spins to a heat bath at temperature $T$.
This Gaussian white noise is characterized by a zero mean $\mean{\vec{\xi}} = 0$ and correlation \cite{Brown63_ThermalFluctuactionsMagnParticles}
\begin{align*}
  \mean{\xi^l_\beta(t) \xi^k_\eta(t')} = \frac{2\alpha \muS \kB T }{\gamma}\delta_{lk}\delta_{\beta\eta}\delta(t-t'),  \\
 \mbox{with}  \quad         l,k\in\{1,...,N\}, \; \beta,\eta\in\{x,y,z\},  
\end{align*}
$\kB$ being the Boltzmann constant.
The $3N$ coupled stochastic differential equations \eqref{eq:LLG} are solved numerically by the stochastic version of Heun's method.
The material parameters define our system of reduced units, $\left|J\right|$ for the energy, $t_J := \nicefrac{\muS}{\gamma \left|J\right|}$ for the time, $a$ for the distance and $B_J := \nicefrac{\left|J\right|}{\muS}$ for magnetic fields.

Our aim is to characterize the noise of the magnetization, which we define here dimensionless as,
\ba{
  \vec{m}(t) = \frac{1}{N}\sum_{n=1}^N \vec{S}^n(t).
}
In thermal equilibrium the mean value $\mean{\vec{m}}$ takes on a finite value in the ferromagnetic phase and vanishes for a paramagnet or for an antiferromagnet even below the critical temperature. 
Furthermore, there is the second moment $\big\langle m_\beta^2 \big\rangle - \mean{m_\beta}^2$ ($\beta\in\{x,y,z\}$), which is zero in the thermodynamic limit. 

Because of that the noise of a macroscopic magnet can hardly be measured, only either a sufficiently small magnet --- a nanoparticle --- or when a small fraction of a sample is probed. 
In spin-noise spectroscopy the latter is the case: the laser is focused on a small area and thus the variance takes on finite values. In our simulations the noise is finite anyhow, as we always have a finite number of spins.

The noise in a steady state can be characterized in two ways: the first is given by the spectral noise power density of the magnetization $P_\beta$, given by the Fourier transform of the magnetization components $\beta\in\{x,y,z\}$,
\begin{align}
  P_\beta(\omega) = \lim_{t^*\to\infty} \frac{1}{t^*} \mean{ \abs{\int_0^{t^*} m_\beta(t) e^{-i\omega t} \; \dd t }^2 },  \label{eq:DefNoisePower}
\end{align}
where $\mean{...}$ denotes the thermal ensemble average.
However, in a steady state this property carries the very same information as the autocorrelation function,
\begin{align}
  \mathrm{Corr}_\beta(t) & = \mean{ m_\beta(t_0 + t) m_\beta(t_0) } = \mean{ m_\beta(t) m_\beta(0) }  \label{eq:DefCorr}  \\
                         & = \frac{1}{2\pi} \int_{-\infty}^\infty P_\beta(\omega) e^{i\omega t} \; \dd \omega,  \label{eq:WKTheorem}
\end{align}
which is the second way.
The former \cref{eq:DefCorr} is independent of $t_0$ in equilibrium or in a steady state.
The latter  \cref{eq:WKTheorem} is the inverse Wiener-Khinchin theorem \cite{Petroni20_StochasticProcesses}, which connects the two properties --- spectral noise power density and correlation function.
Hence, there are always two possible ways to calculate either quantity: either directly or by detour via the other one and use of the (inverse) Wiener-Khinchin theorem. 

In spite of their strict equivalence given by \cref{eq:DefCorr,eq:WKTheorem}, in numerical calculations both ways may not work equally well. In our simulations, firstly, we use the auto\emph{covariance} function instead of the auto\emph{correlation},
\begin{align}
  \mathrm{Cov}_\beta(t) & = \mean{ m_\beta(t) m_\beta(0) } - \mean{m_\beta(t)} \mean{m_\beta(0)}  \label{eq:DefCov}  \\
                        & = \mathrm{Corr}_\beta(t)   - \mean{m_\beta(t)}\mean{m_\beta(0)}.  \nonumber
\end{align}
The Wiener-Khinchin also applies to the autocovariance function for $\omega \neq 0$ since $\mean{m(t)} = \mean{m(0)}$ is time independent and only alters the $\omega = 0$ - mode.
For vanishing mean magnetization (e. g.\ for para- and antiferromagnets without applied magnetic field), autocovariance and autocorrelation coincide anyhow.
It turns out that in the numerical calculations, the former property---the autocovariance---is preferable over the latter as it vanishes in the long-time limit $t\to\infty$ also for finite mean magnetization.
The second practical issue is that it turns out that \cref{eq:DefNoisePower} averages better than \cref{eq:DefCorr} or \cref{eq:DefCov}, when using a finite number of magnetization trajectories. Consequently, it is better to calculate the spectral noise power density first and then the autocovariance as Fourier transform. 

We focus in this work on the magnetization, as this property that can be measured in experiments, for instance by the Faraday rotation or the magneto-optic Kerr effect. However, in simulations also the fluctuation of the \Neel{} vector could readily be calculated but is hard to measure in experiments.

\begin{figure}
	\includegraphics[width=\linewidth]{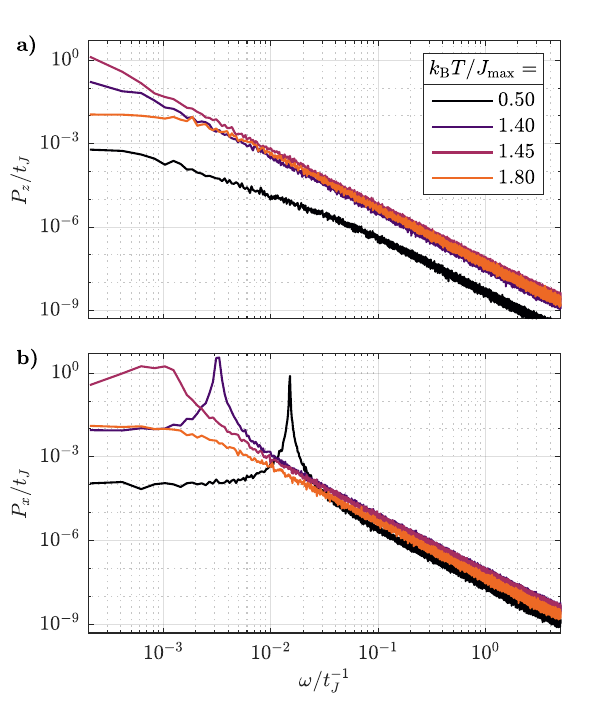}
	\caption{%
	    Spectral noise power density $P_\beta$ of an easy-axis ferromagnet. Upper panel \textbf{a)} shows the noise of $m_z$, lower panel \textbf{b)} of $m_x$ ($m_y$ equivalent to $m_x$).
	    Both exhibit white noise at low frequencies and brown noise at high frequencies.
	    However, the in-plane noise $P_x$ below the Curie temperature $\kB T_\mathrm{C} \approx 1.48\left|J\right|$ has a resonance at the ferromagnetic resonance frequency $\omega_\mathrm{r}$.
	    This frequency decreases with increasing temperature and vanishes right at $T_\mathrm{C}$.
	}
	\label{fig:SpectraFM_tempDep}
\end{figure}

\section{Easy-axis ferromagnets} \label{sec:FM}
We start with the simplest case, an easy-axis ferromagnet, given by $J > 0$ and we use $d_z = 0.01 J$, $\alpha = 0.005$ and $\muS = \mu_{\mathrm{B}}$, which is the Bohr magneton.
For now the applied field remains zero, $\vec{B} = 0$.
We perform $N_\mathrm{av} = 50$ simulations with different realizations of thermal noise, over which the ensemble average is carried out in \cref{eq:DefNoisePower}. 

Because of the uniaxial anisotropy of the model the in-plane components, $x$ and $y$, are equivalent and both differ from the out-of-plane $z$ component.
\Cref{fig:SpectraFM_tempDep} depicts $P_\beta$ for different temperatures, resolved for those two cases.
Above the Curie temperature $\kB T_\mathrm{C} \approx 1.48 J$ all components are qualitatively similar: the low frequencies exhibit white noise $P_\beta = \const$, the high frequencies a brown noise $P_\beta \propto \omega^{-2}$.
Below $T_\mathrm{C}$ a resonance can be found in the in-plane components.
This is a manifestation of a Lorentzian noise power:
\begin{align}
  P_\beta(\omega) = \frac{P^\mathrm{max}_\beta\Gamma_\beta^2}{(\omega - \omega_{0,\beta})^2 + \Gamma_\beta^2} + \frac{P^\mathrm{max}_\beta\Gamma_\beta^2}{(\omega + \omega_{0,\beta})^2 + \Gamma_\beta^2} ,  \label{eq:LorentzianSpectrum}
\end{align}
with central frequency $\omega_{0,\beta}$, half width at half maximum $\Gamma_\beta$ and maximum spectral power density $P_\mathrm{max}$. 
Note that the symmetry $P_\beta(\omega) = P_\beta(-\omega)$ stems from the fact that $m_\beta$ is real valued, and it is not possible to distinguish the sign of any frequency with the setup described here.
A Lorentzian power spectrum  with $\omega_{0,z} = 0$ for the out-of-plane component $P_z$ has also been observed by read heads in the grains of magnetic recording media \cite{Silva90_MagneticNoiseRecordingMedia} for the low frequency range.
By virtue of the inverse Wiener-Khinchin theorem \cref{eq:LorentzianSpectrum} is equivalent to the covariance function
\begin{align}
  \mathrm{Cov}_\beta(t) = \frac{P^\mathrm{max}_\beta}{\tau_\beta} \cos(\omega_{0,\beta}t)\exp\left(\nicefrac{-|t|}{\tau_\beta}\right)  \label{eq:CovarianceFunction}
\end{align}
with the correlation time $\tau_\beta = \nicefrac{1}{\Gamma_\beta}$.

\begin{figure}
	\includegraphics[width=\linewidth]{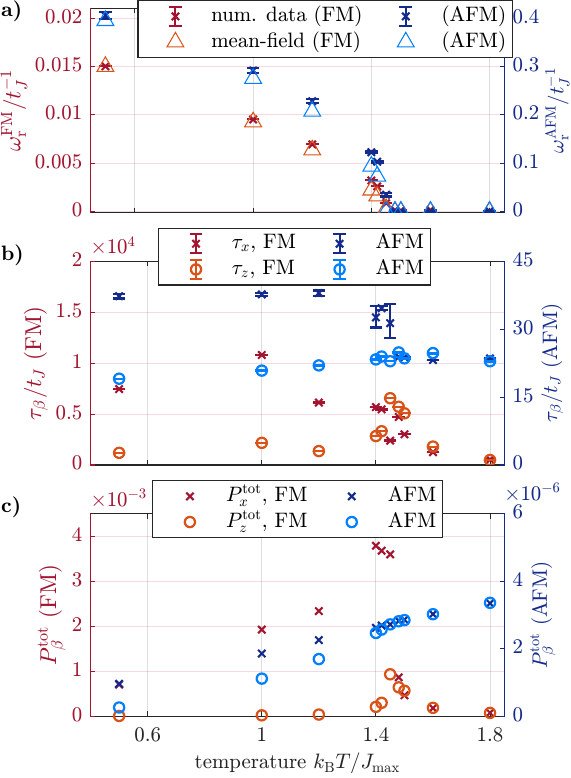}
  \caption{%
    Characteristics for easy-axis ferro- (blue) and antiferromagnets (red) versus temperature $\kB{} T$: 
    \textbf{a)} Resonance frequency $\omega_\mathrm{r}$, numerical data (dark crosses) increasingly deviate from the mean-field estimate \cref{eq:FM_resFrequ} (light triangles), since the mean-field theory gets increasingly inaccurate.
    In either case, $\omega_\mathrm{r}$ vanishes at the critical temperature.
    \textbf{b)} Correlation time $\tau_\beta$ for in-plane $x$- (crosses) and out-of-plane $z$ components (circles). 
    $\tau_z$ peaks for ferromagnets at $T_\mathrm{C}$. Also the ferromagnet exhibits generally much larger values than the antiferromagnet. 
    \textbf{c)} Total noise power $P^\mathrm{tot}_\beta$ for in-plane $x$- (crosses) and out-of-plane $z$ component (circles). 
    For the ferromagnet $P^\mathrm{tot}$ peaks around $T_\mathrm{C}$ for all components. 
    The antiferromagnet rather shows a monotonic increase over the entire temperature range.
  }
  \label{fig:SpectraFM_char}
\end{figure}

In the following we discuss the noise with respect to these three properties, starting with $\omega_0$.
For the $z$ component it is simply zero, but --- most notably --- it takes on finite values below $T_\mathrm{C}$ for the in-plane $x$- and $y$ components and manifests itself as a pronounced peak in the spectral noise power density.
This is attributed to the spin-wave resonance mode, the mode with zero wave vector $\vec{k}=0$.
At zero temperature the spin-wave dispersion is given by \cite{Cramer18_SpinTransportAFM_SSE}
\begin{align*}
  \omega(\vec{k}) \overset{\alpha\ll 1}{=} \frac{\gamma}{\muS}\bigg\{   J\sum_{\mathclap{p=x,y,z}} \left[2 - 2\cos(ak_p)\right] 
                                                                          + 2d_z + \muS B_z \bigg\},
\end{align*}
such that the resonance frequency reads 
\begin{align}
  \omega_{0,xy} \equiv \omega_\mathrm{r} = \frac{\gamma(2d_z+\muS B_z)}{\muS} \label{eq:FM_resonance_0K}.
\end{align}
Here, an applied field $\vec{B}=(0,0,B_z)^\dagger$ is assumed.
\Cref{eq:FM_resonance_0K} is valid if the spins are close to their ground state, $S^l_z \approx 1$, an assumption increasingly inaccurate with increasing temperature.
In a mean-field approximation temperature effects can be included by introducing corresponding micromagnetic parameters, which are temperature dependent ($J \to \mathcal{J}(T)$, $d_z \to \mathcal{D}(T)$, $\muS \to \muS \mean{m_z}$) and replace the atomistic parameters,
\begin{align}
  \omega_\mathrm{r}(T) = \frac{\gamma(2\mathcal{D}(T) + \muS \mean{m_z} B_z)}{\muS \mean{m_z}}. \label{eq:FM_resFrequ}
\end{align}
The temperature dependence in mean-field approximation can be described using the temperature dependence of the order parameter, here the magnetization, via $\mathcal{J} = J \mean{m_z}^2$ and $\mathcal{D} = d_z \mean{m_z}^3$ \cite{Rozsa17_TempDependenceDMI}.
\Cref{fig:SpectraFM_char} a) shows that the numerical data follow this behavior rather well.
Overall we conclude that the phase transition can be measured by means of the temperature-dependence of this resonance peak, which is possible for setups sensitive to in-plane components.

The second property is the half width at half maximum of the resonance peak $\Gamma_\beta$---equivalent to the correlation time $\tau_\beta$.
It has a non-trivial temperature dependence, as can be seen in \cref{fig:SpectraFM_char} b).
For the in-plane component it increases with temperature, reaches its maximum below $T_\mathrm{C}$ and then decreases again. 
The out-of-plane component exhibits the maximum of the correlation time at $T_\mathrm{C}$.
In fact as $m_z$ is the order parameter, it diverges here for an infinite system.
Furthermore, $\tau_z$ scales non-monotonically below $T_\mathrm{C}$.
In the setup here, the correlation time of the in-plane component is larger than the out-of-plane component, even at $T_\mathrm{C}$. However, above $T_\mathrm{C}$ both coincide.
The correlation time also has a distinctive dependence on the magnetic damping: it is easy to imagine that an increased damping decreases $\tau_\beta$, and that in analogy to spin-wave life times it scales antiproportionally with the damping, $\tau_\beta \propto \nicefrac{1}{\alpha}$.
We observe this expected dependence also numerically, see \cref{app:alphaDependence}.

Finally, we show the maximum spectral noise power, $P^\mathrm{max}_\beta$, which --- assuming a Lorentzian --- can also be expressed by the total noise power $P^\mathrm{tot}_\beta$, defined and calculated as 
\begin{align*}
  P^\mathrm{tot}_\beta = 2\int_{\omega > 0} P_\beta(\omega)\,\dd\omega,
\end{align*}
where $P^\mathrm{tot}_\beta = 2\pi P^\mathrm{max}_\beta\Gamma_\beta$ holds true.
\Cref{fig:SpectraFM_char} c) depicts the temperature dependence of the total noise.
This property peaks for both --- the in-plane- and the out-of-plane component --- at the critical temperature and also allows to determine the phase transition.

We want to conclude this section by noting again that the fluctuations can equivalently be characterized by the autocovariance function \cref{eq:CovarianceFunction}.
In an experiment as sketched above, rather this property is measured directly.
The $z$ component follows an exponential decay without oscillation, given by $\omega_0 = 0$.
Components $\mathrm{Cov}_x$ and $\mathrm{Cov}_y$ have an additional oscillation of frequency $\omega_\mathrm{r}$ on top of the exponential decay with correlation time $\tau_\beta$. Moreover, the total noise power $P^{\mathrm{tot}}_\beta$ is reflected in $\mathrm{Cov}_\beta(t=0)=\nicefrac{P^{\mathrm{tot}}_\beta}{2\pi}$.

\begin{figure}
  \includegraphics[width=\linewidth]{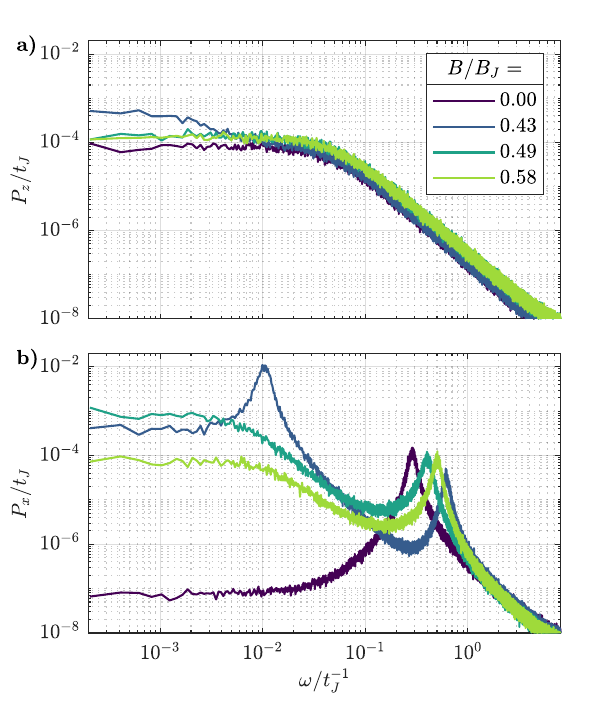}
  \caption{
    Spectral noise power density $P_\beta$ of an easy-axis antiferromagnet for different magnetic fields $B_z$ at fixed temperature $k_\mathrm{B}T = 1\left|J\right|$ below and above the spin-flop field $B_\mathrm{sf} \approx \num{0.45}B_J$ (see \cref{app:MCSpinFlop}).
    \textbf{a)} out-of-plane $z$ component, \textbf{b)} in-plane $x$ component ($P_y$ equivalent). 
    The latter shows the antiferromagnetic resonance frequencies $\omega_\mathrm{r}^\pm$ that have a distinctive field-dependence.
    For fields $0 < B_z < B_\mathrm{sf}$ there are two peaks at finite frequency corresponding to the two non-degenerate magnon branches, above $B_\mathrm{sf}$ one peak is at finite frequency (quadratic branch), the second at zero (linear branch).
  }
  \label{fig:SpectraAFM_fieldDep}
\end{figure}

\section{Easy-axis antiferromagnets}  \label{sec:AFM}
We turn our attention now to antiferromagnetic systems, with $J < 0$, and except the sign of the exchange constant all parameters remain the same, which leads to a magnetic order where nearest-neighbor spins orient in opposite direction --- the so-called chessboard order. 
Probing phase transitions in these magnets is more difficult as the order parameter --- the \Neel{} vector $\vec{n}$ --- often  cannot be measured directly and spin-noise spectroscopy may be an interesting method to apply for these magnets.
Furthermore, in addition to the temperature-induced phase transition to the paramagnetic phase one can often also find a field-induced reorientation transition, called spin flop \cite{Anderson64_StatMech_SpinFlop}, where the \Neel{} vector switches at the spin-flop field $B_\mathrm{sf}$ from orientation along the $z$ axis to an in-plane orientation (in the $x$-$y$ plane).
In the following these phase transitions are studied by means of spin noise.
Even though the average magnetization at zero magnetic field vanishes, the fluctuations do not, and thus an experimental examination of the magnetic order via the spin noise is possible.

Qualitatively, the spectral noise power density of antiferromagnets is akin to their ferromagnetic counterparts, see e. g.\ \cref{fig:SpectraAFM_fieldDep}: low frequencies exhibit white, high frequencies brown noise.
Hence, the general shape of the spectra is Lorentzian.
Furthermore, the in-plane components feature sharp resonances.
This allows to investigate this system for the same three characteristics as the ferromagnet, namely the resonance frequencies $\omega_\mathrm{r}$, the total noise power $P^\mathrm{tot}_\beta$ and the correlation time $\tau_\beta$, $\beta\in\{x,y,z\}$, along with their temperature- and field dependence.

The dispersion relation from the linear spin-wave theory reads (for $\alpha \ll 1$) \cite{Cramer18_SpinTransportAFM_SSE}
\begin{align}
    \omega_\pm(\vec{k}) = \frac{\gamma}{\muS}
                      \left\{ \begin{array}{ll} 
                         \muS B_z \pm \sqrt{ (J_0 + 2d_z)^2  - J_{\vec{k}}^2 }, & B_z < B_\mathrm{sf}\\ 
                         \sqrt{ (J_0 \pm J_{\vec{k}})(J_0^B \mp J_{\vec{k}}^B) }, & B_z > B_\mathrm{sf} 
                      \end{array} \right.  ,
    \label{eq:DispAFM}
\end{align}
from which the resonance frequencies $\omega^\pm_{\mathrm{r}} = \abs{ \omega_\pm(0) }$ follow.
$B_\mathrm{sf}$ denotes the spin-flop field, at zero temperature given by $B_\mathrm{sf} = \nicefrac{2\sqrt{d_z(J_0 - d_z)}}{\muS}$.
Abbreviations $J_0$, $J_{\vec{k}}$, $J^B_0$ and $J^B_{\vec{k}}$ are given by \cref{eq:DispAFM_abrrev}.
Again for a rough estimate at finite temperatures, we use the mean-field approximation for the scaling of the micromagnetic parameters, i.e.\ we assume $J\to\mathcal{J}(T)=J\mean{n_z}^2$, $d_z\to\mathcal{D}(T) = d_z\mean{n_z}^3$, $\muS \to \muS \mean{n_z}$. 

We start the discussion with the temperature dependence at zero field $\vec{B} = 0$. 
Qualitatively, the spectra look like the ones for the ferromagnet, therefore the data for $P_\beta(\omega)$ are not shown explicitly (except for temperature $\kB T = 1\left|J\right|$ in \cref{fig:SpectraAFM_fieldDep}, the spectrum for zero field).
The resonance frequency $\omega_\mathrm{r}$ as a function of temperature is depicted in \cref{fig:SpectraFM_char} a).
We observe that it acts as a measure for the phase transition, as it vanishes at the \Neel{} temperature $\kB T_\mathrm{N} \approx \num{1.48}\left|J\right|$.
In addition the numerical values follow the mean-field estimate fairly well.
Turning our attention to the correlation time $\tau_\beta$, we do not expect a divergence, as the magnetization is not the order parameter.
Actually, $\tau_z$ is a bad measure for the phase transition, see \cref{fig:SpectraFM_char} b), as it has only a very weak temperature dependence.
But the in-plane components $\tau_{x,y}$ can be utilized as they exhibit a drop in the vicinity of $T_\mathrm{N}$.
Just as for the ferromagnet, all components $\tau_\beta$ coincide in the paramagnetic phase.
Moreover, the dependence on the magnetic damping $\alpha$ also is the same, $\tau_\beta \propto \nicefrac{1}{\alpha}$ (see \cref{app:alphaDependence} for details). 
The total noise power $P^\mathrm{tot}_\beta$ also hardly allows to determine $T_\mathrm{N}$ (see panel c) of  \cref{fig:SpectraFM_char}) since all components $P^\mathrm{tot}_\beta$ exhibit a monotonic increase with temperature.
But --- if in-plane- and out-of-plane components are measured separately --- the phase transition is marked by the fact that above $T_\mathrm{N}$ all components equal, whereas below the out-of-plane is smaller than the in-plane component.
In summary, the antiferromagnetic-paramagnetic phase transition does not exactly influence the spin noise as in the ferromagnetic case, however, it leaves traces which can be used for a measurement.

\begin{figure}
	\includegraphics[width=\linewidth]{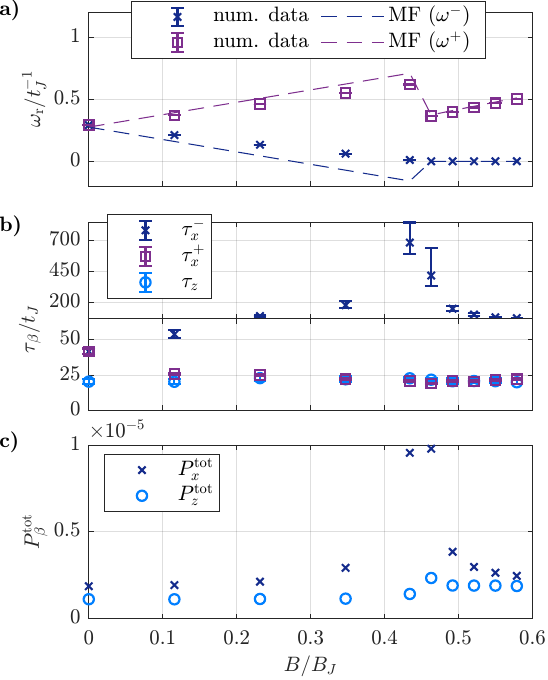}
  \caption{%
    Magnetic field dependence of the noise characteristics of an easy-axis antiferromagnet:
    \textbf{a)} Resonance frequencies $\omega_\mathrm{r}^\pm$ (numerical data as crosses and squares) corresponding to the two magnon branches $\omega^\pm(\vec{k})$ along with a comparison to mean-field estimates (MF, dashed lines).
    For the lower branch, $\omega_\mathrm{r}^-$ vanishes in vicinity of the spin-flop field $B_\mathrm{sf} \approx \num{0.45}B_J$.
    The negative values from the mean-field theory are an artifact of the approximation. 
    For the upper branch, $\omega_\mathrm{r}^+$ jumps at $B_\mathrm{sf}$ but remains finite.
    \textbf{b)} The correlation time $\tau_\beta$ comprises three values, two for the $x$ component according to the two resonances at $\omega^\pm_\mathrm{r}$ (crosses and squares), and a single one for the $z$ component (circles).
    The one for the lower branch $\tau^-_x$ shows a distinctive behavior in the vicinity of the spin-flop transition.
    \textbf{c)} Total noise $P^\mathrm{tot}_\beta$ for the in-plane $x$- (crosses) and out-of-plane $z$ component (circles).
    Both exhibit a maximum at $B_\mathrm{sf}$. 
  }
  \label{fig:SpectraAFM_char}
\end{figure}

We now turn our attention to the magnetic-field dependence, assuming a field along $z$ direction $\vec{B} = B_z\vec{e}_z$, for which we fix the temperature at $\kB T = 1\left|J\right|$.
One might expect a complication in the spin-flop phase since the order parameter has a continuous rotational symmetry in the $x$-$y$ plane and the \Neel{} vector may wander during the simulated time in the plane.
However, it turns out that this does not influence the results, as we discuss in \cref{app:SpinFlopSim} in detail.

To continue with the main investigation, the corresponding field-dependent spectral noise power density is presented in \cref{fig:SpectraAFM_fieldDep}.
Most notably, a finite field lifts the degeneracy of the two spin-wave branches, and the resonance peak splits into two frequencies, corresponding to $\omega^\pm_{\mathrm{r}}$ from \cref{eq:DispAFM}.
As \cref{eq:DispAFM} states, the two resonance frequencies have opposite sign (according to opposite senses of precession). However this sign cannot be resolved in the setup considered here.
Moreover, the spectra $P_{x,y}(\omega)$ form now a sum of two Lorentzians corresponding to the two modes.
The higher resonance frequency $\omega^+_{\mathrm{r}}$ increases with increasing field up to the spin-flop field $B_\mathrm{sf}$, where it jumps to a smaller value and then increases again.
The lower one, $\omega^-_{\mathrm{r}}$, decreases with increasing field until it reaches nearly zero at the spin-flop field $B_\mathrm{sf}$.
In the spin flop phase it remains zero.
This behavior stems from the fact that the spin-wave branches above the spin flop are qualitatively different, see \cref{eq:DispAFM}. There is a gapless linear branch (which shows a linear wave-vector dependence of the frequency for small wave vectors) and a gaped quadratic branch. 
Thus, there is only one branch with a finite gap and the power spectra comprise a single peak at the finite resonance frequency of this branch. 
The gapless linear branch yields a peak at zero frequency in the power spectra.
The behavior of the resonance frequencies along with the mean-field estimate is summarized in \cref{fig:SpectraAFM_char} a).

The behaviour of the correlation time is more complex as compared to the ferromagnet, see panel b) of \cref{fig:SpectraAFM_char}: first of all, for finite field the in-plane components $P_{x,y}(\omega)$ show two correlation times $\tau^\pm_{x,y}$ corresponding to the two peaks in the spectral noise power (above $B_\mathrm{sf}$ one at zero and one at finite frequency).
The lower of the two increases with field up to $B_\mathrm{sf}$, where it peaks, then it falls off again.
The higher one shows only a weak field dependence: it decreases for lower field but then stays approximately constant, even across the spin flop.
The $z$ component shows no significant field dependence at all, however, in the vicinity of the spin flop $P_z(\omega)$ does not exactly follow a Lorentzian, which we address in the last section.
If we consider the total noise power, as depicted in \cref{fig:SpectraAFM_char}, panel c), $P^\mathrm{tot}_\beta$ peaks for all components in the vicinity of the phase transition.

In conclusion, antiferromagnetic order can be probed by spin-noise spectroscopy as well and both phase transitions considered here leave their marks in the noise.

\section{Conclusive discussion}
Within an atomistic spin dynamics approach, we investigate how spin-noise spectroscopy can be utilized to probe magnetic order and explore the features that can be expected in an experiment.
Measuring the time evolution of the magnetization either via Faraday rotation or magneto-optical Kerr effect in a probe-probe setup can give access to the equilibrium fluctuations of the magnetization via its auto-correlation function \cite{Weiss24_SpinNoise_Methods}. From Fourier transformation follows then the spectral noise power density.
In particular, for antiferromagnets, a pump-probe setup reaches the relevant time- and frequency scale, which is terahertz.

Generally, we observe that our numerical spectra are well described by a Lorentzian shape (or a sum of Lorentzians). We identify three key parameters here --- the correlation time, the resonance frequency and the noise amplitude (or alternatively total noise power) --- that capture the main physics and can be utilized to investigate magnetic phase transitions. Measuring an in-plane component, the spectrum exhibits a pronounced peak at the resonance frequency in the ordered phase, which vanishes above the ordering temperature --- clearly marking the critical temperature. But also other properties --- the total noise power or the correlation time --- may signal the phase transition to the paramagnetic phase. Specifically, the total noise power is rather promising as it may peak at a phase transition and therefore gives a strong signal, which reduces the demand for high sensitivity in experiments.
For the antiferromagnet we also detail how the spin-flop transition influences the noise spectra, which highlights that reorientation transitions can be probed as well.
These are rather common for antiferromagnets, besides the spin flop in easy-axis magnets, there is --- for instance --- the Morin transition in hematite \cite{Lebrun20_Hematite_SpinTransport,Dannegger23_HematiteModel} and the reorientation transition in orthoferrites \cite{Tsymbal07_SpinReorientationOrthoferrite,Weiss23_SpinNoise_Orthoferrite}. 

Although the characteristics of a Lorentzian spectral power density, i.e. white noise for low frequencies, a $\nicefrac{1}{\omega^2}$-dependence for high frequencies and a resonance (two resonances for the antiferromagnet with a magnetic field)  are given in all cases, it is not always possible to accurately fit a Lorentzian to the data. The spectra of systems at rather low temperatures and close to the spin-flop transition show deviations from a Lorentzian.
To explore this behavior is beyond the scope of this work and remains a task for the future, but it also highlights that noise correlations carry arguably more information on the magnetic order.

One can also speculate on other aspects that are not covered in our semi-classical approach, one example are quantum fluctuations, which might alter the correlation function at very low temperatures.
In addition in experiments other effects might have an effect, like formation and annihilation of magnetic domains in the vicinity of a phase transition.

Certainly the full potential and versatility of spin-noise spectroscopy for magnetic solids remains to be explored.

\begin{acknowledgments}
  We acknowledge the funding by the Deutsche Forschungsgemeinschaft (DFG, German Research Foundation) via the SFB 1432 under Grant No. 425217212. All simulations were performed on SCCKN, the high-performance computer cluster at the University of Konstanz.
\end{acknowledgments}

\appendix

\section{Damping dependence of the noise-power spectra} \label{app:alphaDependence}
\begin{figure*}
  \includegraphics[width=0.5\linewidth]{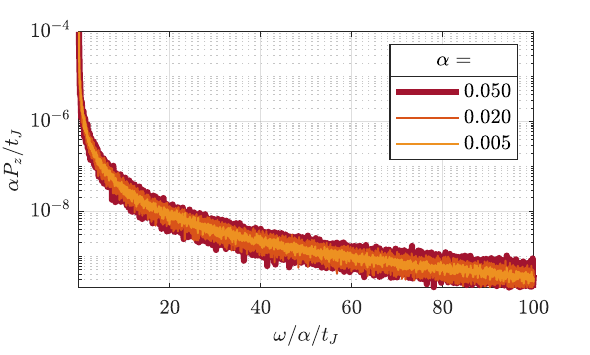}%
  \includegraphics[width=0.5\linewidth]{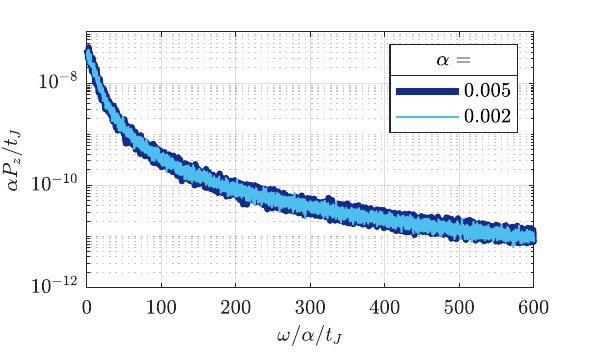}\\%
  \includegraphics[width=0.5\linewidth]{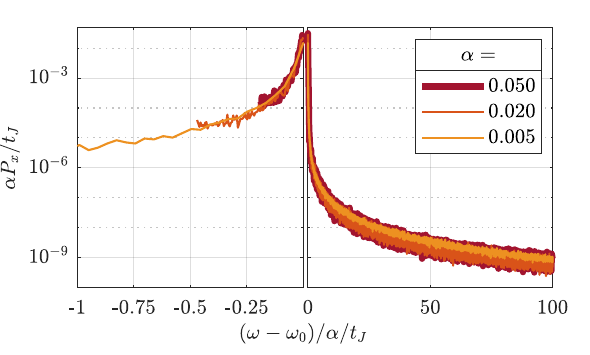}%
  \includegraphics[width=0.5\linewidth]{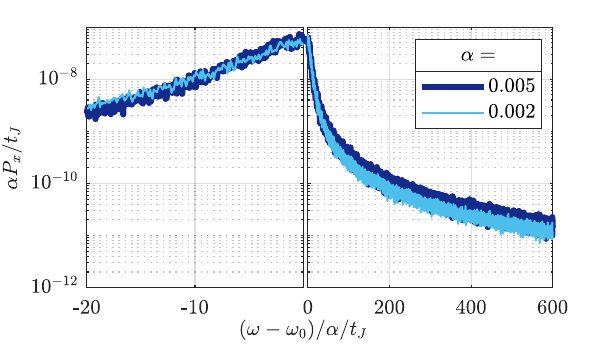}%
  \caption
  {%
    Dependence of the spectral noise power $P_\beta(\omega)$ on the Gilbert damping $\alpha$: we show $\alpha P_\beta$ over $\frac{1}{\alpha}( \omega - \omega_0 )$ for ferromagnets (left panels) and antiferromagnets (right panels) for different values of $\alpha$, where $\omega_0 = \omega^{\mathrm{FM},\mathrm{AFM}}_\mathrm{r}$ is the respective resonance frequency. The temperature is $\kB T = 1\left|J\right|$.
    For each case the spectra coincide for different values of $\alpha$. 
    This proves (within our accuracy) that only $\Gamma_\beta \propto \alpha$ depends on the Gilbert damping.
  }
  \label{fig:alphaDependence}
\end{figure*}

We show numerically that the correlation times $\tau_\beta$ for ferro- and antiferromagnets scale inversely proportional to the Gilbert damping $\alpha$.
For this we assume that the numerical spectral noise power follows \cref{eq:LorentzianSpectrum}.
We restrict the discussion to the positive frequency range, $\omega > 0$, and we define $A^\beta_0 = \mathrm{Cov}_\beta(0) = P_\beta^\mathrm{max}\Gamma_\beta$, which is the amplitude of the covariance function.
The latter property is certainly independent of $\alpha$ since it is (modulo a prefactor) the susceptibility --- an equilibrium property. 
This allows to normalize the spectrum with respect to $\alpha$,
\ba{
  P_\beta(\omega) & = \frac{A_0\Gamma}{(\omega - \omega_0)^2 + \Gamma^2}  \nonumber \\
  \Rightarrow \alpha P_\beta(\omega) & = \frac{A_0 \cdot \left(\frac{\Gamma}{\alpha}\right)}{\left(\frac{\omega}{\alpha} - \frac{\omega_0}{\alpha}\right)^2 + \left(\frac{\Gamma}{\alpha}\right)^2}.
} 
If the hypothesis $\tau_\beta \propto \nicefrac{1}{\alpha}$ is true, then $\Gamma_\beta = \nicefrac{1}{\tau_\beta} \propto \alpha$ and the numerical values for $\alpha P_\beta$ plotted over $\frac{1}{\alpha}(\omega - \omega_0)$ should coincide for different values of $\alpha$.
Exactly this behavior is depicted in \cref{fig:alphaDependence}, showing the evidence for the original assumption $\tau_\beta \propto \nicefrac{1}{\alpha}$ for $P_x$ and $P_z$ for ferro- as well as for antiferromagnets each.

Note here that this argument is only valid for the types of magnets we investigate here --- with easy axis anisotropy --- and for the range of Gilbert damping parameters simulated. Nevertheless, these are representative for experimental values.

\section{Linear spin-wave theory for easy-axis antiferromagnets}
There are different formulations of linear spin-wave theory and the corresponding dispersion relations --- e. g.\ within a full quantum theory or the semi-classical model used in this work \cite{Rezende19_AfmMagnons}. In order to use expressions fitting to our formulation, we concisely recapitulate the calculation of the dispersion relation for antiferromagnets at zero temperature for our atomistic spin model.

We use the following abbreviations
\ba{
  \begin{split}
      J_0           & = N_\mathrm{nb}\abs{J} \\
      J_{\vec{k}}   & = 2\abs{J}\sum_{\mathclap{\substack{p = \\ x,y,z}}} \cos(ak_p) \\
      J_0^B         & = J_0 - 2d_z\left[ 1 - \left(\frac{\muS B_z}{2J_0 - 2d_z}\right)^2 \right] \\
      J_{\vec{k}}^B & = 2\abs{J} \left[ 1 - 2\left(\frac{\muS B_z}{2J_0 - 2d_z}\right)^2 \right] \sum_{\mathclap{\substack{p = \\ x,y,z}}} \cos(ak_p) 
  \end{split} .
  \label{eq:DispAFM_abrrev}
}

The dispersion relation for an easy-axis antiferromagnet (of chessboard order) with field along the $z$ axis---as we consider it in this work --- requires to distinguish two cases with the magnetic field either below or above the spin flop field $B_\mathrm{sf}$.
At zero temperature $B_\mathrm{sf}$, the critical field to switch the antiferromagnet from the easy-axis state ($\vec{S}_l = \pm \vec{e}_z$) to the spin-flop state ($\vec{S}_l = \cos(\vartheta)\vec{e}_z \pm \sin(\vartheta)\vec{e}_x$), can be determined analytically \cite{Anderson64_StatMech_SpinFlop}
\ba{
	B_\mathrm{sf} = \frac{ 2\sqrt{d_z(J_0 - d_z)} }{ \mu_s }.
	\label{eq:AFM_SpinFlopField}
}
In the spin-flop state the angle to the $z$ axis is 
\[
  \cos(\theta) = -\frac{\muS B_z}{2N_\mathrm{nb}J + 2d_z} .
\]

The spin-wave dispersion is given by \cref{eq:DispAFM} and turns out to be different in the easy-axis and the spin-flop state:
Below $B_\mathrm{sf}$ we linearize \cref{eq:LLG} around the easy-axis ground state, which yields 
\ba{
  \omega_\pm(\vec{k}) = \frac{\gamma}{\muS}\left[ \muS B_z \pm \sqrt{(2d_z+J_0)^2 - J_{\vec{k}}^2} \right].
}
for $\alpha \ll 1$ \cite{Cramer18_SpinTransportAFM_SSE}.

In the spin-flop state we first define a transformation that maps the ground state to a colinear state ${^\mathrm{A}\!}\tilde{\vec{S}} = \vec{e}_x$, ${^\mathrm{B}\!}\tilde{\vec{S}} = -\vec{e}_x$.
We use this mapping to transform the LLG equation \eqref{eq:LLG} to the corresponding equation of motion for the mapped spin system and linearize this equation, and Fourier transform it to the resulting linear system:
\ba{
  & \frac{\dd}{\dd t} \begin{pmatrix} {^\mathrm{A}\!}\tilde{S}_y \\ {^\mathrm{A}\!}\tilde{S}_z \\ {^\mathrm{B}\!}\tilde{S}_y \\ {^\mathrm{B}\!}\tilde{S}_z  \end{pmatrix}
       = \frac{-\gamma}{\muS(1 + \alpha^2)} M_{\vec{k}} \cdot 
        \begin{pmatrix} {^\mathrm{A}\!}\tilde{S}_y \\ {^\mathrm{A}\!}\tilde{S}_z \\ {^\mathrm{B}\!}\tilde{S}_y \\ {^\mathrm{B}\!}\tilde{S}_z \end{pmatrix} \\
  & M_{\vec{k}} = \begin{pmatrix}
                  \alpha J_0         &         J_0^B         & \alpha J_{\vec{k}} &        J^B_{\vec{k}} \\
                        -J_0         &  \alpha J_0^B         &       -J_{\vec{k}} & \alpha J^B_{\vec{k}} \\
                  \alpha J_{\vec{k}} &        -J^B_{\vec{k}} & \alpha J_0         &       -J_0^B         \\
                         J_{\vec{k}} &  \alpha J^B_{\vec{k}} &        J_0         & \alpha J_0^B
                \end{pmatrix}  \nonumber .
}
We take the imaginary part of the eigenvalues $\lambda_\pm$ in the limit of small damping $\alpha \ll 1$ and get the eigen frequencies
\ba{
  \omega_\pm(\vec{k}) = \imag{\lambda_\pm} \stackrel{\alpha {\ll} 1}{\approx} \frac{\gamma}{\muS} \sqrt{ (J_0 \pm J_{\vec{k}})(J_0^B \mp J^B_{\vec{k}}) } .
}

\section{Determination of the spin-flop field at finite temperature}\label{app:MCSpinFlop}
\begin{figure}
	\includegraphics[width=\linewidth]{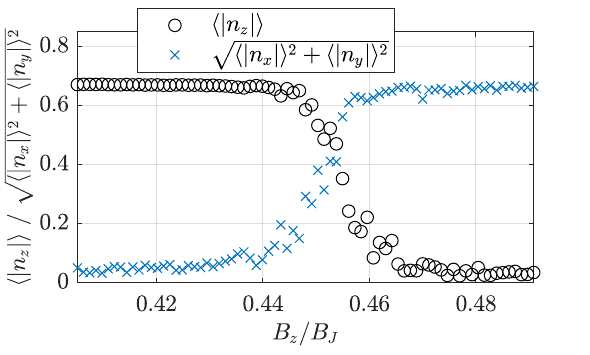}
	\caption{Field dependence of the normalized \Neel{} vector components of the easy-axis antiferromagnet at $k_\mathrm{B}T=1\left|J\right|$ obtained via Monte-Carlo simulations. At $B_\mathrm{sf} \approx 0.45B_J$ the spin-flop transition occurs.}
	\label{fig:AFM_SpinFlopFieldMC}
\end{figure}

The spin-flop field for an easy axis antiferromagnet can be calculated analytically at temperature $T=0$ (see \cref{eq:AFM_SpinFlopField}) and is then given by $B_\mathrm{sf} \approx \num{0,49}B_J$. 
In order to get an approximation of $B_\mathrm{sf}$ at a finite temperature of $k_\mathrm{B}T = 1 \left|J\right|$ we perform Monte-Carlo simulations based on the Metropolis algorithm for the system of the easy-axis antiferromagnet described in \cref{sec:AFM} with a system size of $N = \num{64}^3$. 
The initial configuration of the system is set between the easy-axis and the spin-flop state, being $\vec{S}^{\mathrm{A}/\mathrm{B}}_\mathrm{init} = \pm\nicefrac{1}{\sqrt{2}} \left(1,0,1\right)^\dagger$ with the sign depending on the respective sublattice A and B. 
Note that in our simulation setup there is no well-defined ground state for the flopped antiferromagnet as the orientation of the spins in the $x$-$y$-plane is energetically degenerate, which is discussed in \cref{app:SpinFlopSim}. 
In the Monte-Carlo simulations $\num{50000}$ thermalization steps are performed before the average of the corresponding physical quantity is calculated out of  $\num{100000}$ Monte Carlo steps. 
The so obtained values for the \Neel{} vector $\vec{n}$ are shown in \cref{fig:AFM_SpinFlopFieldMC}. 

At $k_\mathrm{B}T =1\left|J\right|$ the antiferromagnet undergoes the spin-flop transition at an external field of about $\num{0.45}B_J$, which is lower than the spin-flop field at zero temperature but still of comparable order of magnitude.

\section{Simulation setup above the spin-flop field}  \label{app:SpinFlopSim}
Numerical calculation of the spin noise in the spin-flop state comes along with a problem: the \Neel{} vector (the order parameter) has a continuous rotational symmetry in the $x$-$y$ plane.
The magnetic order will therefore wander in the plane. 
In experiments this might not be a problem since sample sizes are so large, that there is barely any such effect noticeable.
Also, in real-world materials, the magnetic order is usually pinned at defects or boundaries.
In contrast, in our simulations the system is rather small and there are no defects nor boundaries (when using periodic boundary conditions). 

\begin{figure}[h]
	\includegraphics[width=\linewidth]{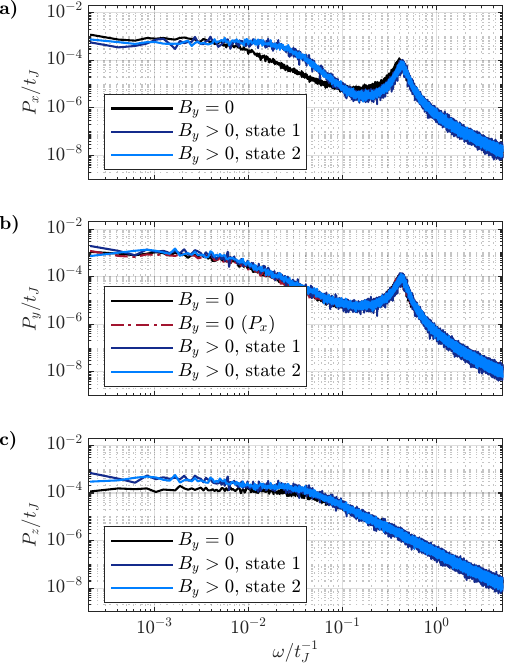}%
	\caption{%
		Comparison of an antiferromagnet in the spin-flop state without finite field ($B_y = 0$) and as reference with $B_y > 0$.
		\textbf{a)} $x$ component of the spectral noise power density $P_x$: with in-plane field the spectrum is notably altered.
		\textbf{b)} $y$ component $P_y$, no significant different with and without in-plane field is observable. Also for comparison also $P_x$ for $B_y = 0$ is depicted to show that $P_y$ for $B_y > 0$ is akin to $P_x$ for $B_y = 0$.
		\textbf{c)} $z$ component $P_z$, there is only a slight difference between the two scenarios for small frequencies $\omega$. 
	}
	\label{fig:SpectraAFM_inplaneField}
\end{figure}

This raises the question if this hampers the calculation of the in-plane noise power density $P_{x,y}$.
To check whether it does, we perform the following numerical test:
We simulate the antiferromagnet in the spin-flop with an additional in-plane field $B_y > 0$, which induces a small magnetization into that very direction (it cants the magnetic order into that direction). This breaks the rotational invariance and locks the magnets ground state in one of two possible states (with \Neel{} vector along $\pm\vec{e}_y$).
The two states complicate the study, as the relaxation process leads the magnetic order into one of the two randomly.
Consequently, we analyze both cases separately (below the data are denoted ``state 1'' and ``state 2'' respectively).
So the total magnetic field reads $\vec{B} = (0, B_y, B_z)^\dagger$ and we choose $B_y$ much smaller than $B_z$. 
We calculate the spectra $P^{B_y}_{x,y}$ and compare them to the original ones $P^0_{x,y}$ (where $B_y = 0$).
The results are depicted in \cref{fig:SpectraAFM_inplaneField}.

For two components, $y$ and $z$, all spectra coincide, which means that the small in-plane field does not change the noise-power notably, regardless of the in-plane orientation of the \Neel{} vector in the $B_y = 0$ case.
In contrast, for the $x$ component the in-plane field does quantitatively alter the spectrum, even though the general shape remains the same. This allows the conclusion, firstly, that for $B_y = 0$ both in-plane components equal for the spin-flop state, regardless what the exact in-plane orientation of the \Neel{} vector is. 
Which in turn shows, secondly, that an in-plane drift of the order parameter does not alter the results notably and the spectra for $B_y = 0$ can readily be analyzed.

\FloatBarrier
\bibliography{Literature} 

\end{document}